\documentclass[letterpaper,english,aps,prb,twocolumn,floatfix,showpacs,amsfonts,amssymb,superscriptaddress]{revtex4}

\usepackage[T1]{fontenc}
\usepackage[latin1]{inputenc}
\usepackage{graphics}
\usepackage{amssymb}
\usepackage{amsmath}
\usepackage{overpic}

\newcommand{\beq}{\begin{equation}}
\newcommand{\eeq}{\end{equation}}
\newcommand{\bea}{\begin{eqnarray}}
\newcommand{\eea}{\end{eqnarray}}

\newcommand{\rmd}{{\rm d}}
\newcommand{\rmi}{{\rm i}}

\newcommand{\cuoo}{CuO$_{2}$}
\newcommand{\lco}{La$_{2}$CuO$_{4}$}
\newcommand{\lno}{La$_{2}$NiO$_{4}$}

\begin{document}

\title{Magnetic susceptibility anisotropies in a
  two-dimensional quantum Heisenberg antiferromagnet
  with Dzyaloshinskii-Moriya interactions}

\author{M.~B.~Silva~Neto}

\email{barbosa@phys.uu.nl}

\affiliation {Institute for Theoretical Physics, University of Utrecht,
  P.O. Box 80.195, 3508 TD, Utrecht, The Netherlands}

\author{L.~Benfatto}

\affiliation
{SMC-INFM and Department of Physics, University of Rome ``La
  Sapienza'',\\ Piazzale Aldo Moro 5, 00185, Rome, Italy}

\author{V.~Juricic}

\affiliation {Institute for Theoretical Physics, University of Utrecht,
  P.O. Box 80.195, 3508 TD, Utrecht, The Netherlands}

\author{C.~Morais~Smith}

\affiliation {Institute for Theoretical Physics, University of Utrecht,
  P.O. Box 80.195, 3508 TD, Utrecht, The Netherlands}

\date{\today}

\begin{abstract}

The magnetic and thermodynamic properties of the two-dimensional (2D)
quantum Heisenberg antiferromagnet (QHAF) that incorporates both 
a Dzyaloshinskii-Moriya (DM) and pseudo-dipolar (XY) 
interactions are studied within the framework of a generalized
nonlinear sigma model (NLSM). We calculate the static uniform 
susceptibility and sublattice magnetization as a function of 
temperature and we show that: i) the magnetic-response is anisotropic
and differs qualitatively from the expected behavior of a conventional 
easy-axis QHAF; ii) the N\'eel second-order phase transition becomes 
a crossover, for a magnetic field $B\perp$ {\cuoo} layers. We provide 
a simple and clear explanation for all the recently reported 
{\it unusual} magnetic anisotropies in the low-field susceptibility 
of {\lco}, L.~N.~Lavrov {\it et al.}, Phys. Rev. Lett. {\bf 87}, 
017007 (2001), and we demonstrate explicitly why {\lco} can not be 
classified as an ordinary easy-axis antiferromagnet.

\end{abstract}


\maketitle

\section{Introduction}

The usual starting point for the description of the magnetism in
undoped cuprates is the 2D isotropic Heisenberg model
\beq
H=J\sum_{<ij>}{\bf
S}_i\cdot{\bf S}_j,
\eeq
with only the AF super-exchange $J$ between the nearest-neighboring
Cu$^{++}$ spins, ${\bf S}_i$.\cite{kastner} Remarkably, a long-wavelength 
NLSM description of such model has been shown to correctly capture many of 
the features observed experimentally in the high-temperature paramagnetic
phase.\cite{CHN, CSY} At lower temperatures, however, small anisotropies 
like the DM and XY interactions become important, and the commonly
used isotropic description is no longer adequate to describe the
long-wavelength magnetic properties of {\lco}. In a recent paper, 
Lavrov {\it et al.} reported on some {\it unusual} anisotropies in the 
low-field magnetic susceptibility observed in detwinned {\lco} single 
crystals.\cite{Ando-Mag-Anisotropy} Among the unexpected features one 
could mention: i) an almost featureless transverse in-plane 
susceptibility, $\chi_{a}$, for an extended temperature range; ii) 
the increase of the longitudinal susceptibility, $\chi_b$, as one 
approaches the N\'eel transition temperature, $T_{N}$, from the 
ordered side; iii) the anomalous $T=0$ hierarchy 
$\chi_c>\chi_{b}>\chi_{a}$, where $a,b,c$ represent the 
low-temperature orthorhombic (LTO) crystallographic directions of 
Fig.\ \ref{Fig-1}. 

In this article we investigate theoretically the magnetic response and
thermodynamic properties of the square-lattice QHAF that incorporates both
the DM and XY anisotropy terms by investigating the generalized NLSM
corresponding to this microscopic problem. As it has been stressed in the
context of one-dimensional systems,\cite{Affleck} the DM term generates an
effective staggered magnetic field proportional to the applied uniform
field.  In the two-dimensional case the continuum field theory allows one
to reproduce straightforwardly the same effect, \cite{Papanicolaou,sces05}. 
In Ref.  \cite{Papanicolaou} the attention was focused on the (classical) 
spin configurations expected in an external field. In this paper we 
investigate instead the thermodynamical properties of the system, which 
enter in a crucial way in the evaluation of the static uniform susceptibility, 
$\chi$, as a function of temperature, $T$.  In particular, we show that 
the coupling induced by the DM term between the magnetic field and the 
antiferromagnetic order parameter is particularly relevant when the 
order-parameter fluctuations become critical, leading to an anomalous 
magnetic response of the system, as observed for example in
\cite{Ando-Mag-Anisotropy}. As an outcome of our studies we observe that,
while the isotropic nonlinear sigma model (NLSM) is well established to
give a proper description of the magnetism in the high-temperature
paramagnetic phase, $T\gg T_{N}$,\cite{CHN,CSY} for the ordered (broken)
phase, $T<T_{N}$, we find more appropriate the adoption of a {\it  soft} 
version of such fixed-length constraint, in order to correctly capture
the physics of the longitudinal fluctuations.  For the sake of
clarity, we refer explicitly to the {\lco} system throughout the paper, 
but the results presented here apply to any 2D square-lattice QHAF where 
the DM and/or XY anisotropy interactions are present, as for example 
other cuprates and nickelates {\lno}.\cite{Shekhtman}

\section{The long-wavelength limit}

\begin{figure}[htb]
\includegraphics[scale=0.4]{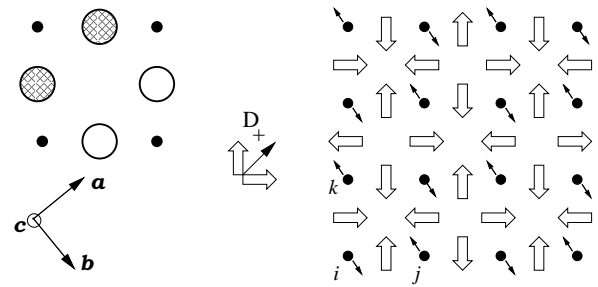}
\caption{Left: the hatched circles represent the O$^{--}$ ions
  tilted above the {\cuoo} plane; the empty ones are tilted below
  it; small black circles are Cu$^{++}$ ions; $bac$ orthorhombic coordinate
  system. Right: Schematic arrangement of the staggered magnetization
  (small black arrows) and DM vectors (open arrows). Center: continuum
  definition of ${\bf D}_{+}$.}
\label{Fig-1}
\end{figure}

We consider the following square-lattice single-layer $S=1/2$ 
Hamiltonian for the {\lco} system
\beq
H=J\sum_{\langle i,j\rangle}{\bf S}_{i}\cdot{\bf S}_{j}+
\sum_{\langle i,j\rangle}{\bf D}_{ij}\cdot\left({\bf S}_{i}\times{\bf
    S}_{j}\right)+
\sum_{\langle i,j\rangle}{\bf S}_{i}
\cdot\overleftrightarrow{\bf \Gamma}_{ij}\cdot{\bf S}_{j},
\label{Hamiltonian}
\eeq
where ${\bf D}_{ij}$ and $\overleftrightarrow{\bf \Gamma}_{ij}$ are,
respectively, the DM and XY anisotropic interaction terms that arise due to
the spin-orbit coupling and direct-exchange in the low-temperature
orthorhombic (LTO) phase of {\lco}.\cite{Shekhtman} The direction and the
alternating pattern of the DM vectors, shown in Fig.\ \ref{Fig-1}, are a
direct consequence of the tilting structure of the oxygen octahedra and of
the symmetry of the {\lco} crystal.  Throughout this work we adopt the LTO
$(bac)$ coordinate system of Fig.\ \ref{Fig-1}, for both the spin and
lattice degrees of freedom, and we use units where $\hbar=k_{B}=1$. Thus we
have that 
\beq
{\bf D}_{ij}=\frac{1}{\sqrt{2}}(-d,d,0), {\;\;\;\;\;\;\;\;}
{\bf D}_{ik}=\frac{1}{\sqrt{2}}(d,d,0), 
\eeq
and
\bea
\label{eq:Gamma}
\overleftrightarrow{\Gamma}_{ij}{=}
  \left( \begin{array}{ccc}
     \Gamma_1 & \Gamma_2 & 0 \\
     \Gamma_2 & \Gamma_1 & 0 \\
     0 & 0 & \Gamma_3
         \end{array} \right)\!\!,\;
\overleftrightarrow{\Gamma}_{ik}{=}
  \left( \begin{array}{ccc}
     \Gamma_1 & -\Gamma_2 & 0 \\
     -\Gamma_2 & \Gamma_1 & 0 \\
     0 & 0 & \Gamma_3
         \end{array} \right)\!\!,
\nonumber
\eea
where $ij$ and $ik$ label the Cu$^{++}$ sites (see Fig.\ \ref{Fig-1}), 
and $d$ and $\Gamma_{1,2,3}>0$ are of order $10^{-2}$ and $10^{-4}$,
respectively, in units of $J$.\cite{Shekhtman}

To construct the long wavelength effective theory for the above
Hamiltonian we follow the standard procedure: we write
\beq
\frac{{\bf S}_i(\tau)}{S}=e^{\rmi{\bf Q \cdot x}_i}
{\bf n}({\bf x}_i,\tau)+ {\bf L}({\bf x}_i,\tau),
\eeq
where ${\bf n}$ and ${\bf L}$ are, respectively, the staggered and
uniform components of the spin and ${\bf Q}=(\pi,\pi)$. We then 
integrate out ${\bf L}$ and we obtain a modified NLSM where 
additional terms appear due to the DM and XY interactions ($\beta=1/T$)
and 
\begin{widetext}
\beq
{\cal S}=\frac{1}{2 g_0 c_0}\int_{0}^{\beta}\rmd\tau\int\rmd^{2}{\bf x}
\left\{(\partial_{\tau}{\bf n})^{2}+
 c_0^2(\nabla{\bf n})^{2}+({\bf D_+}\cdot {\bf n})^2+
\Gamma_c\; n_c^2 \right\}, 
\label{nlsm}
\eeq 
\end{widetext}
and the fixed length constraint ${\bf n}^2=1$ is implicit (see also
Ref. \cite{Papanicolaou,sces05}).  Here $g_0$ is the bare coupling constant, 
related to the spin-wave velocity, $c_0$, and renormalized stiffness, 
$\rho_s$, through $\rho_{s}=c_0(1/Ng_0-\Lambda/4\pi)$,\cite{CSY,CHN} 
where $\Lambda$ is a cutoff for momentum integrals (we set the lattice 
spacing $a=1$) and $N=3$ is the number of spin components. Finally,
\beq
{\bf D}_+=2S({\bf D}_{ij}+{\bf D}_{ik}),
\eeq
see Fig.\ \ref{Fig-1}, and
\beq
\Gamma_c=32JS^2(\Gamma_1-\Gamma_3)>0,
\eeq
and we neglected a small anisotropic correction to $c$ coming 
from the XY term, since $\Gamma_{1,2}\ll J$. 

%
%
%

Observe that in the NLSM (\ref{nlsm}) only the sum of the DM and XY 
terms on the $(ij)$ and $(ik)$ bonds appears. This has several 
consequences: (i) the XY term generates a contribution to 
(\ref{nlsm}) proportional to $-\Gamma_{\alpha\alpha} n_\alpha^2$, 
$\alpha=a,b,c$, where $\Gamma_{\alpha\beta}=(1/2)(\Gamma_{ij}+
\Gamma_{ik})_{\alpha\beta}=\mbox{diag}(\Gamma_1, \Gamma_1,\Gamma_3)$. 
Since ${\bf n}^2=1$, such contribution reduces to a constant shift in 
the classical energy plus the $\Gamma_c{\;}n_c^2$ term in Eq. (\ref{nlsm}). 
This allows us to identify the $ab$ as the first easy plane. 
The property that only the combination $\Gamma_1-\Gamma_3$ affects the 
physics of the model (\ref{Hamiltonian}) was an outcome of the numerical 
results of the RPA improved spin-wave analysis of Ref. \cite{Gooding}, but no
explanation for this effect was provided. (ii) The DM distortion 
enters the action (\ref{nlsm}) only through the vector 
${\bf D_+}=D_+ \hat{\bf x}_a$ ($\hat{\bf x}_a,\hat{\bf x}_b$,
and $\hat{\bf x}_c$ are the LTO unit vectors), rendering $bc$ a 
second easy plane. When the system orders we find the staggered 
magnetization at $\langle{\bf n}\rangle=\sigma_0\hat{\bf x}_b$, 
as observed experimentally,\cite{Vaknin} because the orientation 
along $\hat{\bf x}_a$ or $\hat{\bf x}_c$ would cost an energy $D_+^2$ or 
$\Gamma_c$, respectively. Moreover, since the uniform 
magnetization is given by \cite{Papanicolaou,sces05} 
\beq
\langle{\bf L}\rangle=\frac{1}{2J}(\langle{\bf n}\rangle\times{\bf D}_{+}),
\eeq
we find that, in the AF phase, $\langle{\bf L}\rangle$, is directed
along $\hat{\bf x}_c$, so that the Cu$^{++}$ spins are canted 
out of the $ab$ plane and confined to the $bc$ plane. 

\section{Magnetic Susceptibility}

To evaluate the static uniform susceptibility, $\chi_\alpha$, we
rederive from the microscopic Hamiltonian (\ref{Hamiltonian})
the quantum action (\ref{nlsm}) in the presence of a magnetic field 
${\bf B}$ (in units of $g_S\mu_B/\hbar=1$, where $g_S\approx 2$ 
is the gyromagnetic ratio and $\mu_{B}$ is the Bohr 
magneton)\cite{Papanicolaou,sces05}
\bea
{\cal S}({\bf B})&=&
{\cal S}(\partial_{\tau}{\bf n}\rightarrow\partial_{\tau}{\bf n}+
\rmi{\bf B}\times{\bf n}) \nonumber\\
&+&\frac{1}{g_0c_0}
\int_{0}^{\beta}\rmd\tau\int\rmd^{2}{\bf x} {\;} {\bf B}\cdot({\bf
  D}_{+}\times{\bf n}).
\label{Action-Mag-Field}
\eea
The last term in the above equation, which couples the DM vector ${\bf
D}_{+}$ and the staggered magnetization ${\bf n}$, is responsible for many
of the {\it unusual} features observed experimentally in {\lco}. However,
this coupling is generic to any square-lattice system, as other cuprates
and nickelates, where symmetry guarantees that the DM vectors alternate 
in sign between neighboring bonds.\cite{Shekhtman}

The {\it zero-field} magnetic susceptibility is calculated in linear
response theory as 
\beq
\chi_{\alpha}=\left.\frac{1}{\beta V}\frac{\partial^{2}\log{Z}}{\partial
B_{\alpha}^{2}}\right|_{B=0},
\eeq
where $Z({\bf B})=\int{\cal D}{\bf n}\exp\left\{-{\cal S}
({\bf B})\right\}$ is the Euclidean partition function for the 
action (\ref{Action-Mag-Field}), and we obtain
\bea
\chi_a&=&\chi_{a}^{u}+\frac{\sigma_0^2}{g_0c_0},\nonumber\\
\chi_b&=&\chi_{b}^{u}+\frac{D_+^2}{g_0c_0}G_c(0,0),\nonumber\\
\chi_c&=&\chi_{c}^{u}+\frac{\sigma^2_0}{g_0c_0}+\frac{D_+^2}{g_0c_0}G_b(0,0),
\label{Susceptibilities}
\eea
where 
\beq
G^{-1}_{\alpha}({\bf k},\omega_{n})=c_0^{2}{\bf k}^{2}+\omega_{n}^{2}+
M_{\alpha}^{2}
\eeq
is the inverse propagator for the magnetic modes $a,b,c$, with 
gaps $M_{\alpha}$ and
\beq
\chi_{\alpha}^{u}=\frac{1}{\beta V}\sum_{q=({\bf k},\omega_{n})}
\left\{G_{\beta}(q)+ G_{\gamma}(q)-
4\omega_n^2 \; G_{\beta}(q)\; G_{\gamma}(q)\right\},
\eeq
is the traditional ``uniform'' contribution to the susceptibility, with
$(\alpha\beta\gamma)=(abc)$ and its permutations. 

The set of Eqs.\ (\ref{Susceptibilities}), following from the quantum
actions (\ref{nlsm}) and (\ref{Action-Mag-Field}), is the main results of 
this article. A few remarks are in order now concerning the usefulness
of the NLSM approach to the model (\ref{Hamiltonian}) with respect to 
other approaches.\cite{Shekhtman,Thio,Gooding} First, the action 
(\ref{Action-Mag-Field}) allows one to easily track down the source 
of the anisotropic magnetic response, as coming from the term 
${\bf B} \cdot ({\bf D}_+\times {\bf n})$. In fact, depending on the
directions of ${\bf D}_+$ and of the applied field, there will be (or not)
additional terms proportional to $G_\alpha$ in Eq.\ (\ref{Susceptibilities}), 
responsible for the {\it unusual} magnetic response. Second, this 
coupling is generic to any system where the lack of inversion-center
symmetry allows for an oscillating DM interaction between neighboring 
spins. Third, Eqs.\ (\ref{Susceptibilities}) are 
{\it exact}, within linear response theory, even though the evaluation 
of the zero-field thermodynamic quantities $\sigma_0(T)$ and 
$M_{\alpha}(T)$ will require, in general, a certain degree of 
approximation. Nevertheless, we show now that all the qualitative 
features observed in \cite{Ando-Mag-Anisotropy} are already present 
at the mean-field level.

The thermodynamic properties of the model (\ref{nlsm}) can be 
investigated by means of a large-$N$ expansion.\cite{CSY} At 
$N=\infty$, and for $T>T_N$, we find $\sigma_0=0$ and
\beq
M_{b}^{2}=c_0^2\xi^{-2},\;
M_{a}^2=D_+^2+c_0^2\xi^{-2},\; M_{c}^2=\Gamma_c+c_0^2\xi^{-2},
\label{equations-disordered}
\eeq
where the correlation length, $\xi$, is calculated through an 
{\it averaged} constraint equation $1=NI_\perp(\xi)$. Here
$I_\perp=(1/2)(I_{a}+I_{c})$ with
\beq
I_{\alpha}= \frac{g_0 c_0}{\beta V} \sum_{{\bf k},\omega_{n}} G_{\alpha}({\bf k},\omega_{n})= 
\frac{g_0T}{2\pi c_0}
\log\left\{\frac{\sinh\left(c_0\Lambda/2T\right)}
{\sinh\left(M_{\alpha}/2T\right)}\right\}.
\label{prop}
\eeq
Since both transverse spin-wave modes, $a$ and $c$, are gapped, the 
2D system orders at a finite N\'eel temperature, $T_N$, defined by 
$1=NI_\perp(\xi=\infty)$.\cite{J-perp} 

For the ordered phase, $T<T_N$, instead of replacing the local constraint 
by an {\it average} one, we introduce a potential of the type 
$(u_{0}/2)({\bf n}^{2}-1)^{2}$ in the action (\ref{nlsm}), making 
the constraint locally {\it soft}. In this case we obtain a generalized
{\em linear} sigma model (LSM), which leads to an ordered phase described by:
\beq
\sigma_{0}^{2}=1-NI_{\perp},{\;}
M_{b}^{2}=2\sigma_{0}^{2}u_{0},{\;}
M_{a}^{2}=D_+^2,{\;} M_{c}^{2}=\Gamma_c.
\label{equations-ordered}
\eeq
We should emphasize that the only new parameter here is the energy 
scale $u_0=(\gamma J)^2$, where $\gamma$ is a fitting parameter that
will be fixed later through comparison with experiments. However, an 
inspection of Eq. (\ref{equations-ordered}), indicates that neither 
$T_N$ nor the values of $\sigma_{0}$ and $M_{a,c}$ depend on $u_0$.
The only difference is the physics of the longitudinal fluctuations, 
for $T<T_{N}$. While the hard constraint yields a longitudinal mode 
always gapless,\cite{Katanin} the soft one gives a mass for the 
longitudinal mode below $T_N$ that is proportional to the strength 
of the order parameter. 

%
%
\begin{figure}[htb]
\includegraphics[scale=0.35,angle=-90]{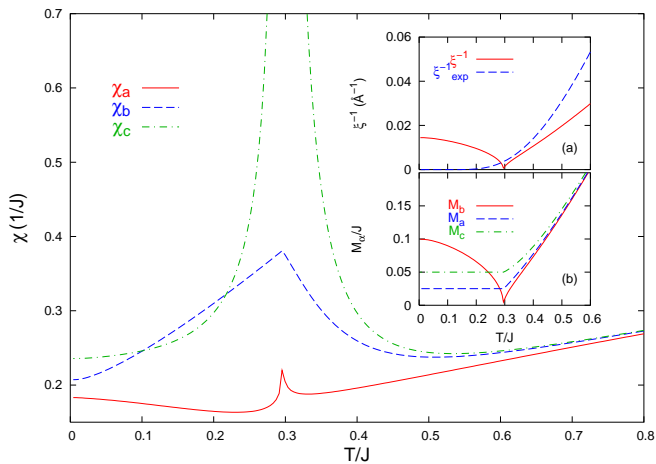}
\caption{(Color online) Temperature dependence of the magnetic 
susceptibilities  $\chi_{a}$, $\chi_{b}$ and $\chi_{c}$ from 
(\ref{Susceptibilities}). Insets: (a) temperature
dependence of $M_{a,b,c}$ and (b) $\xi^{-1}$.}
\label{Fig-2}
\end{figure}
%

We fix the large-$N$ parameters for {\lco} by comparing the correlation length,
$\xi$, with the experimental data in the paramagnetic phase, and we 
use $M_{a}(T=0)\equiv D_+=2.5$ meV and $M_{c}(T=0)\equiv 
\sqrt{\Gamma_c}=5.0$ meV, for the zone center in-plane and
out-of-plane spin-wave gaps as determined from neutron 
scattering.\cite{Keimer} In the inset (a) of Fig.\ \ref{Fig-2} we compare 
$\xi^{-1}$ obtained for $\rho_s=0.1 J$, $c_0=1.3J$ and $J=100$ meV with 
the curve $\xi_{exp}$, which represents the best fit of the experimental 
data at $T>T_N$ obtained in the isotropic NLSM\cite{Keimer2} (so that 
$\xi_{exp}^{-1}$ deviates from the experimental data near $T_N$). When 
$1/N$ corrections to our calculations are included one expects that the
ratio $T_N/J$ is reduced,\cite{Katanin} which allows for larger
and more realistic values for $\rho_s$, $c_0$, and $J$.\cite{CHN,CSY} 
Nevertheless, with the above parameters we find $\sigma_{0}(T=0)\approx 0.46$
which leads to an effective moment of the Cu$^{++}$ spins,
$\mu=g_SS\sigma_{0}\mu_{B}\approx 0.46\mu_{B}$, in good
agreement with the experiments.\cite{Keimer2} The resulting
$M_{a,b,c}$ as a function of temperature are shown in the inset (b) 
of Fig.\ \ref{Fig-2} for $\gamma=0.15$, a choice that will 
be justified below.

\subsection{Temperature dependence of $\chi_a,\chi_b$ and $\chi_c$}

Let us now discuss how all the unusual features of the susceptibilities
reported in\cite{Ando-Mag-Anisotropy} follow straightforwardly from Eqs.
(\ref{Susceptibilities}) already at the mean-field (large N) level (see
Fig.\ \ref{Fig-2}). At large temperatures, $T\gg T_N$, all three
$\chi_{a,b,c}$ exhibit the usual linear-$T$ behavior observed
experimentally\cite{Nakano} and expected for the uniform susceptibility of
the isotropic NLSM.\cite{CSY} As $T\rightarrow 0$, on the other hand,
$\chi^u_b\rightarrow 0$ while both $\chi_c^u$ and $\chi_a^u$ saturate at
very small, but nonzero, values. The smooth evolution through the
transition experimentally observed in $\chi_a$ is due to the fact that 
$\chi_{a}^u$ contains terms like $I_{\alpha}$ in Eq.\ (\ref{prop}),
which are regular for the transverse modes (always gapped by the DM 
and XY interactions) and diverge logarithmically at $T_N$ for the 
longitudinal mode as $M_{b}(T\rightarrow T_N)\rightarrow 0$. This
divergence appears as a hump in the numerical evaluation of $\chi_a$ 
presented in Fig.\ \ref{Fig-2} and is an artifact of both the low 
dimensionality and of the large-$N$ limit (a coupling between {\cuoo} 
layers or a nonzero anomalous dimension $\eta$ for the propagator of 
the longitudinal mode regulates such logarithmic infrared divergence 
of $I_{b}$ at $T_N$ and wipes out the feature in $\chi_a$). Finally, 
the weak increase observed in $\chi_a$ as $T\rightarrow 0$, also 
observed experimentally, is due to the compensation between the 
decrease of $\chi^u_a$ and the increase of the term $\sigma_0^2/g_0c_0$.

We turn now to $\chi_c$ and $\chi_b$. As $T\rightarrow T_N$, 
\beq
\chi_c\approx \frac{D_+^2}{g_0 c_0} G_b(0,0)=
\frac{D_+^2}{g_0c_0}\frac{1}{M_{b}^2}
\eeq
diverges because of the vanishing of the mass of the longitudinal 
mode, $M_{b}(T\rightarrow T_N^+)\rightarrow 0$, and this is
associated with the ferromagnetic ordering of the canted
moments.\cite{Thio} It is essential to have a finite $M_{b}$ 
below $T_{N}$ in order to correctly reproduce the behavior of 
$\chi_c$ in the ordered phase. So, although the hard-spin model 
correctly reproduces the vanishing of $M_{b}$ as $T\rightarrow T_{N}^{+}$, 
it is not adequate for the ordered phase, where $M_{b}=0$ leading 
to a diverging $\chi_{c}$ for all $T<T_{N}$. That is why we have
adopted the soft version of the constraint below $T_N$. 

Finally, we observe in Fig.\ \ref{Fig-2} that $\chi_{b}$ is not divergent
at $T_N$ but exhibits a well pronounced peak. This is a consequence of the
term $D_+^2G_c(0,0)/g_0c_0=D_{+}^2/(g_0c_0 M^2_{c})$.  According to
Eq. (\ref{equations-disordered}), as we approach $T_{N}$ from the
paramagnetic side $\xi^{-2}$ goes to zero and $D_{+}^2/(g_0c_0 M^2_{c})$
approaches the value $D_{+}^2/(g_0c_0 \Gamma_{c})\approx 1/(4g_0c_0)$,
which is controlled by the ratio between the DM and XY interaction
terms. Observe that the absence of a similar feature in $\chi_{a}$ is due
to the fact that, for ${\bf B}\parallel a$, ${\bf B}\cdot({\bf
D}_{+}\times{\bf n})=0$, so $\chi_{a}$ has only the uniform contribution
$\chi_{a}^{u}$. Thus, the increase in $\chi_b$ as $T\rightarrow T_{N}^-$ is
clearly a result of the sum of its uniform part, $\chi_b^u$, which is
monotonically increasing with $T$, and the term $D_{+}^2/(g_0c_0 M^2_{c})$,
which is constant below $T_N$ and decreases instead as $\sim\xi^2$ above
$T_N$.

\subsection{Unusual $\chi_a(0)<\chi_b(0)<\chi_c(0)$ hierarchy}

A comment is in order now concerning the hierarchy of the
$T=0$ susceptibilities in Fig.\ \ref{Fig-2}. For an
ordinary easy-axis AF with different in-plane and out-of-plane
transverse gaps, $M_{a}<M_{c}$, it is expected that the
$T=0$ {\it uniform} susceptibilities should satisfy
$\chi_b^u(0)<\chi_a^u(0)<\chi_c^u(0)$, with the lowest value for
the easy axis and with a transverse susceptibility smaller in the direction
of the smaller gap. This is in fact the result that one obtains after
dropping the terms proportional to $D_+^2$ in Eq.\ 
(\ref{Susceptibilities}). However, the unusual coupling between ${\bf B}$ 
and ${\bf n}$ in (\ref{Action-Mag-Field}) gives rise to the terms 
proportional to $D_+^2$ in $\chi_b(0)$ and $\chi_c(0)$, which, in
turn, lead to the following $T=0$ values for the three susceptibilities
\beq
\chi_a\approx\frac{\sigma_0^2}{g_0c_0}, \quad
\chi_b=\frac{1}{g_0c_0}\frac{D_+^2}{M_{c}^2}, \quad
\chi_c\approx\chi_a+\frac{1}{g_0c_0}\frac{D_+^2}{M_{b}^2}, 
\label{limit-values}
\eeq
resulting in the {\it unusual} hierarchy of the zero-temperature values and
rendering {\lco} an example of an unconventional easy-axis antiferromagnet.
Observe that $\chi_c(0)/\chi_a(0)>1$ always, and we use the $T=0$
intercept of $\chi_c$ to fix $\gamma=0.15$. Finally, for our choice of 
parameters the ratio $\chi_b(0)/\chi_a(0)\sim 1.1$, 
in relatively good agreement with the
experiments,\cite{Ando-Mag-Anisotropy} 
but it depends in general on the values of 
the transverse masses.\cite{notagooding}

%
%
\begin{figure}[htb]
\includegraphics[scale=0.35,angle=-90]{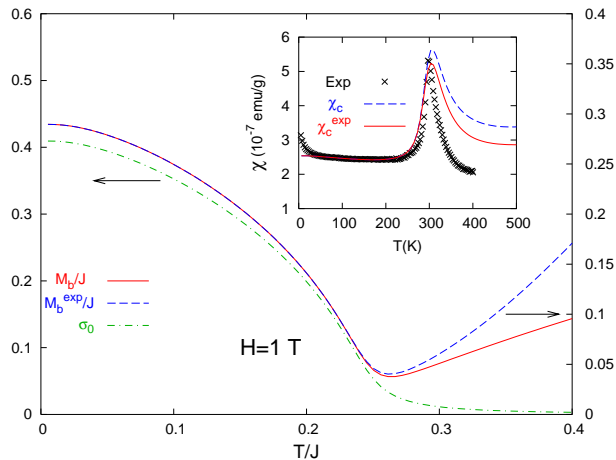}
\caption{(Color online) Plot of $\sigma_0$, $M_{b}$, and $M_b^{exp}$ 
from Eqs.\ (\ref{new-OP}) and (\ref{new-masses}). $M_b^{exp}$ is fitted to
the experimental data for the inverse correlation length above $T_N$, see 
discussion in the text below Eq. (\ref{new-OP}). 
Inset: $\chi_{c}$ in units of emu/g  and experimental points from 
\protect\cite{Ando-Mag-Anisotropy}. $\chi_c^{exp}$ is obtained from 
Eqs. (\ref{Susceptibilities}) using $M_b^{exp}$. Here $\rho_s=0.07J$, 
$\gamma=0.5$, $a=5.3$ {\AA}, the out-of-plane lattice parameter $d=13.5$
{\AA}, and we added a $T$-independent shift of 
$\chi_{\mbox{vV}}\approx 1.0\times 10^{-7}$ emu/g 
to account for the van Vleck paramagnetic 
susceptibility.\cite{Ando-Mag-Anisotropy}}
\label{Fig-3}
\end{figure}
%

\section{Effecs of a finite magnetic field ${\bf B} \perp \mbox{\cuoo}$ layers}

The result for $\chi_c$ plotted in Fig. \ref{Fig-2} can be further improved
by considering explicitly the nontrivial effects of the combination ${\bf
B}\perp ab$ plane and ${\bf D}_+\neq 0$ on the thermodynamic properties of
the theory (\ref{Action-Mag-Field}).  In fact, in this case the last term
in Eq.\ (\ref{Action-Mag-Field}) becomes $-(1/g_0c_0)\int h{\;}n_b$, where
$h=|{\bf B}\times{\bf D}_+|$.  Within the NLSM formulation the averaged
constraint equation defines a self-consistency equation for the order
parameter given by:
\beq
\sigma_0^2=1-NI_\perp(M^2_\perp+h/\sigma_0)
\label{new-OP-NLSM}
\eeq
where the masses appearing in the transverse spin fluctuations (\ref{prop}) are
themselves a function of $\sigma_0$:
\beq
M_a^2=D_+^2+\frac{h}{\sigma_0}, \;
M_c^2=\Gamma_c+\frac{h}{\sigma_0}+B^2. \;
\label{new-masses}
\eeq
As one can see, besieds a (quantitatively small) hardening of the $c$ mode
due to the applied field, we find that the term $h/\sigma_0$ plays a a role
similar to the correlation length, $c^2\xi^{-2}$, in Eq.
(\ref{equations-disordered}). The result is that $\sigma_0$ never vanishes
for $h\neq 0$ (see Fig.\ \ref{Fig-3}), and because of the DM interaction, a
nonvanishing ${\bf B}\perp ab$ plane transforms the N\'eel second order
phase transition into a crossover. In analogy with the discussion of the
case ${\bf B}=0$, the proper description of the longitudinal-field
fluctuations below the crossover temperature can be done within the LSM,
where the order-parameter equation and the longitudinal mass read:
\beq
u_0\sigma_0(\sigma_0^2-1+NI_\perp(M^2_\perp+h/\sigma_0,T))=h, \; 
M_b^2=2u_0\sigma_0^2+\frac{h}{\sigma_0}.
\label{new-OP}
\eeq
Observe that the crossover temperature obtained from Eq.\
(\ref{new-OP-NLSM}) and Eq.\ (\ref{new-OP}) is almost the same. To
interpolate between the two solutions we define
$(M^{exp}_b)^2=max(h/\sigma_0,2u_0\sigma_0^2+h/\sigma_0)$ and we fix the
parameter values by fitting $M_b^{exp}$ to the the experimental data on the
correlation length above $T_N$. As far as the $c$-axis susceptibility is
concerned it is now evident from the inset of Fig.\ \ref{Fig-3} that
$\chi_c$ no longer diverges ($M_b$ is always finite) but becomes peaked at
a crossover temperature at which $M_b$ has a minimum (see Fig.\
\ref{Fig-3}). Observe also that for $\chi_a$ and $\chi_c$ the effects of a
finite magnetic field are instead negligible, leading only to small
quantitative differences with respect to the calculation presented before
for ${\bf B}=0$. 

\section{Conclusions}

In conclusion, we have derived the uniform magnetic susceptibility within
the long-wavelength effective theory (\ref{nlsm})-(\ref{Action-Mag-Field})
for the single-layer square-lattice QHAF with DM and XY interactions
(\ref{Hamiltonian}).  We obtained that the magnetic response is
anisotropic, in remarkable agreement with the experiments of
\cite{Ando-Mag-Anisotropy}, and differs from the expected behavior for a
more conventional easy-axis QHAF. Due to the presence of the DM term, the
uniform magnetic field generates an effective staggered field proportional
to both the DM interaction and to the applied field. We showed that the
coupling between the magnetic field and the staggered order parameter is
particularly relevant when the order-parameter fluctuations become
critical, leading to an anomalous magnetic response of the system, as
observed for example in \cite{Ando-Mag-Anisotropy}. Moreover, this same
coupling is responsible for the unusual zero temperature hierachy of the
susceptibilities, rendering {\lco} an example of an unconventional
easy-axis antiferromagnet. Finally, we considered the effect of a finite
magnetic field on the $\chi_c$ susceptibility. We found that, for a nonzero
${\bf B}\perp ab$ plane, the N\'eel second order phase transition becomes a
crossover, see Fig.\ \ref{Fig-3}, in analogy with the beahvior of a
ferromagnet in a finite magnetic field. The finite-field effects are
instead irrelevant as far as the susceptibilities in the $a$ and $b$
direction are considered. However, the case of ${\bf B}\parallel b$ presents
interesting outcomes as far as the selection rules for one-magnon Raman
scattering are concerned, as it has been discussed recently in
Ref. \cite{Raman}.  While Eq.\ (\ref{Action-Mag-Field}) and
(\ref{Susceptibilities}) have been derived for the case of {\lco}
materials, the present analysis can be extended to other QHAF systems where
the lack of inversion-center symmetry allows for an oscillating DM
interaction between neighboring spins.

\section{Acknowledgements}

The authors have benefitted from discussions with A.~Aharony, Y.~Ando,
B.~Binz, C.~Castellani, A.~H.~Castro~Neto, and R.~Gooding. 
We also thank A.~Lavrov, for fruitfull discussions and for providing
us with the experimental data for $\chi_c$ from \cite{Ando-Mag-Anisotropy},
shown in Fig.\ \ref{Fig-3} (inset).


\begin{thebibliography}{20}
\bibitem{J-perp}
The interlayer coupling, $J_\perp$, plays also an important role 
for the N\'eel transition in {\lco}. We will nevertheless proceed 
with our 2D model since it already captures the essential physics 
reported in \cite{Ando-Mag-Anisotropy}.
\end{thebibliography}
\end{document}